\documentstyle[psfig,proceedings,numreferences]{crckapb} 

\begin{opening}
\title{ENTRAINMENT CONTROL OF CHAOS \protect\\
NEAR UNSTABLE PERIODIC ORBITS
}

\author{R. METTIN}
\institute{Institut f\"ur Angewandte Physik\\
           Technische Hochschule Darmstadt\\
           Schlo\ss gartenstr. 7\\
           64289 Darmstadt, Germany\footnote{Current address:
	   3.\,Phys.\,Inst., Univ. G\"ottingen, B\"urgerstr.\,42-44,
	   D-37073 G\"ottingen}
}

\end{opening}

\runningtitle{ENTRAINMENT CONTROL OF CHAOS NEAR UPOS}

\begin{document}

\begin{abstract}
It is demonstrated that improved entrainment control of chaotic systems
can maintain periodic goal dynamics near unstable periodic orbits
without feedback. The method is based on the optimization of goal
trajectories and leads to small open-loop control forces.

\end{abstract}

\section{Introduction}

In recent years, a large number of investigations has been made
into the control of chaotic dynamics, and many techniques
have been applied in simulations and experiments (see, for instance,
\cite{overview} for an overview).

Two techniques have been proposed for an {\it open-loop}
control 
of chaos. The first approach is related to vibrational methods, as a
scalar periodic perturbation is applied to the chaotic
system. Usually, the control signals are sinusoidal \cite{sinus} or
two-mode forces \cite{twomode}. Recently it has been shown that
the method can be improved by use of optimized multimode signals which
are more complex \cite{mettin2}.

The second method uses equations of motion and a specific goal
dynamics to derive vector control forces. If the goal trajectory is
suitably chosen, the chaotic system under control converges to the goal.
Therefore, this method is often referred to as entrainment
control \cite{entrainment}. In this paper it is shown how entrainment
control can be improved with respect to small control forces. For this
purpose, a specific property of chaotic systems is exploited: dense
unstable periodic orbits (UPOs).

In fact, UPOs are common goal orbits of most feedback techniques
employed for control of chaos (see, e.g., \cite{ott}). Because a UPO
is a natural, but 
unstable motion of the system, control forces have to be applied
only for transfer to and stabilization of the orbit. In the ideal
(noiseless) case, the stabilization forces tend to zero once the UPO is
actually reached. Therefore, feedback control of UPOs can be maintained
with very small forces in low noise systems. Despite the power of such
methods, however, there are situations where one wants to or has to
dispense with a feedback from the system. Then, open-loop techniques
are needed.

While there exists a theory for open-loop stabilization of unstable
fixed points (vibrational control, see \cite{bellman}), a counterpart
for stabilization of unstable periodic orbits is still lacking to the
author's knowledge. Although it is supposed that the scalar periodic
perturbation methods mentioned above usually stabilize periodic dynamics
in the vicinity of UPOs, this has not been shown yet 
explicitly. The underlying mechanism, possibly some resonance
phenomenon, as well as the exact final dynamics are still unknown.
This is different for entrainment control, where the appropriate
dynamics is given {\it a priori}, and which is described in
the following.

\section{Entrainment control}

We start with a nonlinear dynamical system in the continuous time
domain, which is influenced by a control signal vector. The equations
of motion are supposed to be known, according to the ordinary
differential equation 
\begin{equation} \label{e1}
\frac{d{\bf x}(t)}{dt} = \dot{\bf x}(t)={\bf f}({\bf x}(t),{\bf u}(t))
\end{equation}
with the system state ${\bf x}$ in the state space ${\rm I \! R}^n$,
the control vector ${\bf u}$ in the control signal space ${\rm I \!
R}^s$ and the vector field ${\bf f}: {\rm I \! R}^n \times {\rm I \! R}^s
\to {\rm I \! R}^n$. Let ${\bf z}$ be a goal dynamics generated by a
vector field ${\bf g}$, 
$\dot{\bf z}(t) \! = \! {\bf g}({\bf z}(t))$, ${\bf z}(t_0)\! =\! {\bf z}_0$.
To introduce the goal dynamics as solutions of Eq.~(\ref{e1}), we
have to apply forces ${\bf u}(t)$ which solve the equation
\begin{equation} \label{e3}
\dot{\bf x}(t) = {\bf f}({\bf x}(t),{\bf
u}(t)) = {\bf g}({\bf x}(t)).
\end{equation}
Thus, the vector field ${\bf f}$ is simply changed to ${\bf g}$ by the
control forces. To do this, however, feedback from the system is
necessary, as the actual system state ${\bf x}(t)$ appears in
Eq.~(\ref{e3}). The main point of the entrainment control method
is to eliminate ${\bf x}(t)$ by the assumption that the system is
already located in the initial goal state ${\bf z}_0$ when the control
is started at $t_0$. If so, the correct control signal ${\bf u}(t)$ is
given by the solution of the equation
\begin{equation} \label{e4}
{\bf f}({\bf z}(t),{\bf
u}(t)) = {\bf g}({\bf z}(t)) = \dot{\bf z}(t).
\end{equation}
Equation~(\ref{e4}) can be solved without any system state
measurement; in fact, not even a generating vector field ${\bf g}$ has to
be given: it is sufficient to know the goal trajectory itself and its
time derivative (velocity) 
for the control time interval, $\{{\bf z}(t),\dot{\bf
z}(t)\}_{t\in [t_0,t_1]}$.

There are several conditions that have to be fulfilled to make the
entrainment control scheme work. First of all, Eq.~(\ref{e4}) has to
be solvable for ${\bf u}(t)$. This is trivially true for
simple vector additive forces:
\begin{equation} \label{e5}
{\bf f}({\bf z}(t),{\bf u}(t)) =
\tilde{\bf f}({\bf z}(t))+{\bf u}(t) \;\; \Rightarrow \;\; {\bf u}(t) =
\dot{\bf z}(t)-\tilde{\bf f}({\bf z}(t))
\end{equation}
We restrict our
discussion to such forces, which are the most common in the
literature on entrainment control. However, problems immediately arise
if, e.g., the control space dimension $s$ is less than the state
space dimension $n$ (a treatment of general control influence with
$s$=$d$ can be found in \cite{mettin1}).

The next condition to be satisfied is asymptotic stability of the goal
trajectory. While control forces according to Eq.~(\ref{e4}) ensure
that the goal trajectory is a solution of the controlled system
Eq.~(\ref{e1}), there is no statement about whether nearby
located system states are attracted by it -- in other words, whether
entrainment occurs in a vicinity of ${\bf z}(t)$ according to 
$\lim_{t\to \infty} |{\bf x}(t)-{\bf z}(t)|\! =\! 0$.
Even if this is the case, one needs a large basin of attraction
(ideally the whole phase space) to make the method work for a large
set of possible initial states distant from ${\bf z}(t)$. Statements
about basins are very hard to find (compare \cite{jackson}), but a
discussion of stability can more easily be made. For
simple vector additive control, Eq.~(\ref{e5}), we call all points in
phase space where all eigenvalues of the Jacobian of $\tilde{\bf f}$
have negative real part, {\it convergent regions} (see
also \cite{mettin1,jackson}). Goal trajectories entirely located in
convergent regions turn out to be asymptotically stable, if their
time derivative $\dot{\bf z}(t)$ is sufficiently bounded.

\section{Optimization of the goal trajectory}

Because of the stated stability aspects, goal trajectories are
usually chosen to be located in convergent regions. This results in a
typical drawback of the method: Due to very little overlap of convergent
regions and unperturbed chaotic attractor, such a goal dynamics is
quite different to the natural system dynamics. Consequently, the
system is strongly altered by control, and control forces are
large. In fact, they have to be of about the magnitude of the
velocities appearing in the uncontrolled system in order to pull the
movement into convergent regions.

To attack this problem, one has to realize that location in convergent
regions is not a necessary condition for stability of a
goal trajectory. The resulting dynamics of chaotic systems controlled by
periodic perturbation methods indeed suggest that stable
dynamics can also be achieved in the chaotic attractor region,
especially near a UPO. Control forces are smaller then, as
the natural dynamics is only slightly altered. An extreme case would
be to consider a UPO itself as a goal for entrainment control: we get
zero control forces according to Eq.~(\ref{e4}). However, such a goal
is of course unstable. It has to be at least slightly changed to
result in a stable one. To this end, a
family of deformations of a UPO is considered in the following.

Let ${\bf z}_{_{UPO}}(t)$ denote a known UPO of the chaotic system with
period $T$=$2\pi / \omega$. We chose a finite Fourier series as a
deformation $\Delta {\bf z}(t)$, and also include a linear
time transformation by a factor $\eta$. The family of goal
trajectories now reads
\begin{eqnarray} \label{e7}
{\bf z}(t) & = & {\bf z}_{_{UPO}}(\eta t) + \Delta {\bf z}(\eta t)
\nonumber \\
 & = & {\bf z}_{_{UPO}}(\eta t) + \sum_{m=0}^M \left( {\bf a}^m
\cos(m \omega \eta t) + {\bf b}^m \sin(m \omega \eta t) \right).
\end{eqnarray}
It is parametrized by $\eta \in {\rm I \! R}$ and the Fourier
coefficient vectors ${\bf a}^m$, ${\bf b}^m \in {\rm I \!
R}^n$. Since ${\bf b}^0$ has no effect on ${\bf z}(t)$,
a deformation (or a goal) is
characterized by  a total of $d=(2M \! + \! 1)n+1$ real numbers.

Now, we formulate the determination of advantageous deformation
parameters that lead to a stable goal trajectory with small control
forces as an
optimization problem. A real number according to a cost function is
assigned to each probed set of parameters. The cost assesses the stability 
of the chosen goal trajectory (which is determined numerically) as
well as the magnitude of the resulting control forces:
\begin{equation} \label{e8}
cost = \left\{ \begin{array}{l@{\quad:\quad}l}
\|{\bf u}(t)\|_{max} + \gamma [\exp(\mu)-1] & 
\mu < 1, \\
\|{\bf u}(t)\|_{max} + \gamma [\exp(\mu)-1] + C
& \mu \ge 1 \end{array} \right.
\end{equation}
Here, $\mu \! =\! \max_i \{ |\mu_i| \}$ is the maximum absolute value of the
characteristic (Floquet) multipliers of the goal orbit. These are well
defined, as the goal is periodic, and they are calculated by
integration of the variational equations \cite{parker}. Instability is
indicated by $\mu > 1$ and causes high cost via a large positive
penalty term $C$. The cost function further includes an $\exp(\mu)$
and the maximum norm of the resulting forces. The
weight of stability with respect to magnitude of forces can be
adjusted by $\gamma$. The global
minimum of the cost function in the deformation parameter space
corresponds to the best goal trajectory in sense of the chosen balance
between stability and small forces. 

For various reasons, a direct analytic treatment of the given
optimization problem is usually not possible: the UPO is not known in
analytic form, a direct expression for stability of a deformed UPO is
missing (the variational equations have to be integrated), and the
cost function is not continuous. Consequently, numerical methods are
employed. The UPO is represented by a periodic cubic spline
interpolation, and $\mu$ is calculated via numerical integration
of the variational equations. The optimization is
done by a numerical technique that can handle
high-dimensional problems with rough and rapidly varying cost functions.
For this purpose, the stochastically guided algorithm {\tt amebsa}
from \cite{press} was chosen; this is a combination of simulated
annealing and the downhill simplex method.

\begin{figure}[bht]
\centerline{\psfig{figure=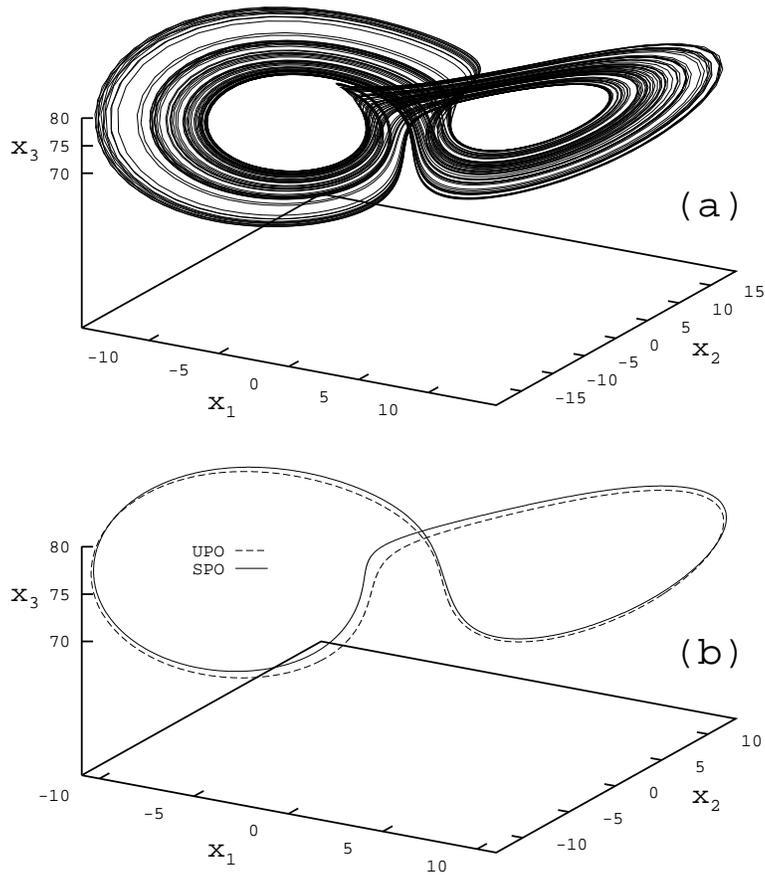,bbllx=80pt,bblly=79pt,bburx=501pt,bbury=568pt,width=10.cm}}
\caption{\label{fig1}(a): Chaotic attractor of the Lorenz system
($\sigma=10.0$, $r=75.0$, $b=0.4$). (b): Embedded unstable periodic
orbit (UPO, interrupted line) and the optimized deformation of it that
turns out to be a globally stable goal trajectory (SPO, solid line).
}
\end{figure}

\pagebreak

\begin{figure}[t]
\centerline{\psfig{figure=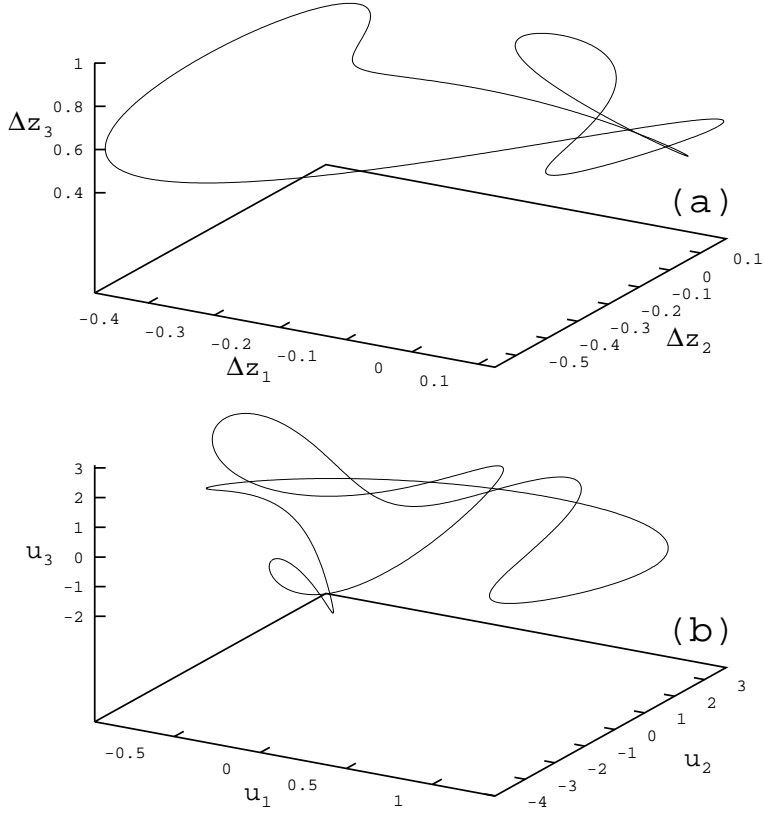,bbllx=70pt,bblly=83pt,bburx=506pt,bbury=558pt,width=10.cm}}
\caption{\label{fig2}(a): Orbit in phase space of the difference
vector $\Delta {\bf z}$ between the UPO of the Lorenz system and the
optimized stable goal trajectory (SPO).
(b): Path of the corresponding vector control forces in the
control signal space. 
}
\end{figure}

\section{Example}

In this section, the control of a chaotic Lorenz system is
demonstrated. The equations with vector additive
control read
\begin{eqnarray} \label{e9}
&&\dot x_1  =  \sigma (x_2 - x_1) + u_1(t),\quad
\dot x_2  =  r x_1 - x_2 - x_1 x_3 + u_2(t), \nonumber \\
&&\dot x_3  =  x_1 x_2 - b x_3 + u_3(t),
\end{eqnarray}
where time dependence of the forces is explicitly written.
A given periodic goal trajectory $\{ {\bf z}(t), \dot{\bf z}(t)\}_{t\in
[0,T]}$ yields control forces according to Eq.~(\ref{e5}).
Parameters of the Lorenz system are set to $\sigma=10$, $r=75$, and
$b=0.4$. The resulting chaotic attractor is shown in
Fig.~\ref{fig1}(a). An embedded UPO which corresponds to ${\bf
z}_{_{UPO}}$ in Eq.~(\ref{e7}) is given in Fig.~\ref{fig1}(b) by the
interrupted line.

The deformation is defined by five Fourier modes ($M$=5) which leads to a
34-dimensional search space for optimization; $\gamma$ is set to 5,
$C$ to 100 in Eq.~(\ref{e8}). The best result of several optimization
runs is shown in Fig.~\ref{fig1}(b) by the solid line. It is a stable
goal trajectory and therefore a stable periodic orbit (SPO) of the
controlled system. The actual values of deformation parameters can be
found in Tab.~\ref{tab1} together with additional data of the UPO and
the SPO. The deformation lies in the range of some percent, and it is
plotted in Fig.~\ref{fig2}(a). The resulting forces, shown in
Fig.~\ref{fig2}(b), change the vector field of the chaotic system
less than about 10\%. This is an improvement of more than
a magnitude if compared to goals in convergent regions \cite{jackson}.

\begin{figure}[htb]
\centerline{\psfig{figure=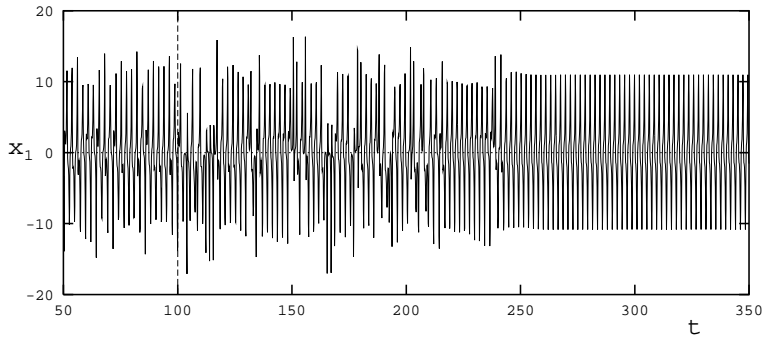,bbllx=63pt,bblly=35pt,bburx=545pt,bbury=252pt,width=10.cm}}
\caption{\label{fig3}Controlled Lorenz system, first coordinate $x_1(t)$:
Control is turned on at $t_0=100$, and the goal trajectory is reached
after transients at about $t=250$ (after approximately~75 control periods).
}
\end{figure}

Numerical tests indicated that the SPO is globally asymptotically
stable; the basin is the whole phase space.
However, transient times until control is established depend strongly
on the initial state, and range from just a few up to a few hundred
control periods. A typical behavior is presented in
Fig.~\ref{fig3}. After control is turned on, an intermittent transient
appears. Finally, the system settles down on the desired goal orbit,
which is maintained.

\section{Conclusion}

It has been shown how open-loop entrainment control in the vicinity of
unstable periodic orbits can be realized. The search for suitable
goal trajectories has been formulated in terms of an optimization
problem with respect to UPO deformations. Feasibility has been
demonstrated in an example, where a chaotic Lorenz system has been
successfully controlled to an optimized distortion of a UPO. The locations
of such goal orbits are independent of convergent regions, and thus the
required forces are small compared to hitherto used goal
dynamics far from the chaotic attractor.

This work was supported by the Deutsche Forschungsgemeinschaft
\linebreak (Sonderforschungsbereich 185).

\begin{table}[htb]
\begin{center}
\caption{\label{tab1} Coordinates $z_{1,2,3}$ give a point of the
original UPO of the Lorenz system, $T$ its period. Maximum absolute
values of the Floquet multipliers are given by $\mu$ for the UPO and for
the optimized deformation (SPO). Parameters of the
SPO are the time transformation coefficient
$\eta$ and Fourier coefficients $a^m_n$, $b^m_n$.
Subscripts of numbers indicate a decimal shift, i.e., $1.234_{(-2)}$
stands for $1.234 \times 10^{-2}$.}
\begin{tabular}{llll}
\hline
$z_{1}=8.0$ & $z_{2}=11.59618$ & $z_{3}=68.87129$ & 
$T=2.153957$\\
\hline
$\mu_{_{UPO}}=2.555$ & $a^0_1=-9.794703_{-2}$ &
$a^1_1=2.303546_{-2}$ & $b^1_1=-1.946829_{-1}$\\
$\mu_{_{SPO}}=0.719$ & $a^0_2=-1.126072_{-1}$ &
$a^1_2=-4.441229_{-2}$ & $b^1_2=-1.543789_{-1}$\\
$\eta=1.001267$   & $a^0_3=6.741204_{-1}$ &
$a^1_3=-9.082832_{-2}$ & $b^1_3=5.702340_{-2}$\\
\hline
$a^2_1=1.063004_{-1}$ & $b^2_1=2.904178_{-5}$ & 
$a^3_1=5.343983_{-2}$ & $b^3_1=-9.527600_{-3}$\\
$a^2_2=1.553288_{-1}$ & $b^2_2=-4.403841_{-2}$ &
$a^3_2=6.312158_{-2}$ & $b^3_2=-2.940390_{-2}$\\
$a^2_3=-2.747584_{-1}$ & $b^2_3=-7.345061_{-2}$ &
$a^3_3=9.633147_{-2}$ & $b^3_3=8.743705_{-2}$\\
\hline
$a^4_1=2.634346_{-2}$ & $b^4_1=8.791274_{-2}$ &
$a^5_1=2.069052_{-3}$ & $b^5_1=1.429860_{-2}$\\
$a^4_2=9.267735_{-2}$ & $b^4_2=6.849559_{-2}$ &
$a^5_2=2.874568_{-2}$ & $b^5_2=2.781748_{-2}$\\
$a^4_3=7.009472_{-2}$ & $b^4_3=2.443292_{-2}$ &
$a^5_3=-2.478029_{-2}$ & $b^5_3=3.759263_{-2}$\\
\hline
\end{tabular}
\end{center}
\end{table}

\end{document}